\documentclass[twocolumn,aps,prd,superscriptaddress,longbibliography]{revtex4-1}

\usepackage{graphicx}
\usepackage{dcolumn}
\usepackage{bm}
\usepackage{braket}
\usepackage{textcomp}
\usepackage{float}
\usepackage{lineno}

\usepackage{soul,color} 

\soulregister\ref{7}
\soulregister\eqref{7}
\soulregister\cite{7}
\soulregister\onlinecite{7}

\begin{document}

\preprint{APS/123-QED}

\title{Customizing the angular memory effect for scattering media}

\author{Hasan Y{\i}lmaz}
\affiliation{Department of Applied Physics, Yale University, New Haven, Connecticut 06520, USA}%
\affiliation{Institute of Materials Science and Nanotechnology, National Nanotechnology Research Center (UNAM), Bilkent University, 06800 Ankara, Turkey}
 \author{Matthias K\"{u}hmayer}
 \affiliation{Institute for Theoretical Physics, Vienna University of Technology (TU Wien), A-1040 Vienna, Austria}%
 \author{Chia Wei Hsu}
\affiliation{Ming Hsieh Department of Electrical and Computer Engineering, University of Southern California, Los Angeles, California 90089, USA}
 \author{Stefan Rotter}
 \affiliation{Institute for Theoretical Physics, Vienna University of Technology (TU Wien), A-1040 Vienna, Austria}%
\author{Hui Cao}
 \email{hui.cao@yale.edu}
 \affiliation{Department of Applied Physics, Yale University, New Haven, Connecticut 06520, USA}%


\begin{abstract}
The memory effect in disordered systems is a key physical phenomenon that has been employed for optical imaging, metrology, and communication through opaque media. Under the conventional memory effect, when the incident beam is tilted slightly, the transmitted pattern tilts in the same direction. However, the ``memory'' is limited in its angular range and tilt direction. Here, we present a general approach to customize the memory effect by introducing an angular memory operator. Its eigenstates possess perfect correlation for tilt angles and directions that can be arbitrarily chosen separately for the incident and transmitted waves, and can be readily realized with wavefront shaping. This work reveals the power of wavefront shaping in creating any desired memory for applications of classical and quantum waves in complex systems.

\end{abstract}

\pacs{Valid PACS appear here}

\maketitle

\section{Introduction}
Multiple scattering of light in disordered media such as white paint, paper, and biological tissue randomizes the propagation of waves and scrambles the spatial information carried by an incident beam. Once the thickness of a sample exceeds the transport mean free path, the information about the incident direction is lost, and light waves are scattered in all directions. While the interference of these scattered waves forms a random grainy pattern (speckle) in transmission, some memory is retained, as a result of the deterministic scattering process~\cite{Akkermans, 1994_BerkovitsPR, 1998_Freund1001}. One prominent example is the angular memory effect: if the incident wavefront of a coherent beam is tilted by a small angle, the transmitted wavefront is tilted by the same amount in the same direction. Both classical and quantum waves possess such a memory~\cite{1994_Genack_PRE, 2015_Gigan_OE, 2017_Vellekoop_Optica, 2018_Defienne_PRL, 2020_Lib_SA}, which persists even in the deep diffusive regime where forward-scattered waves are negligible and the information of the original propagation direction is lost already~\cite{1988_Feng_PRL, 1988_Stone_PRL, 1989_Berkovits_PRB, 1989_Berkovits_PRB2}. In this way the angular memory effect provides unique access to the transmitted far-field pattern behind a disordered sample, which can be conveniently scanned by tilting the incident wavefront. This feature has enabled a wide range of applications in imaging, sensing, and optical metrology through turbid media~\cite{2012_Mosk_Review, 2015_Vellekoop_Review, 2017_Rotter_Gigan_Review, 2020_Choi_Review, 2010_Vellekoop_OL, 2010_Psaltis_OE, 2011_VanPutten_PRL, 2012_Bertolotti_Nat, 2014_Fleischer_OE, 2014_Katz_NatPhoton, 2015_Yilmaz_Optica, 2017_Yang_OE, 2017_Guillon_OL, 2017_Papadopoulos_NP, 2018_Bertolotti_OE, 2018_Waller_Optica, 2018_Katz_OL, 2019_Katz_OL, 2019_Daniel_OE}.

While the angular memory effect exists for any incident wavefront, it is severely limited by the small angular range of $\lambda/(2\pi L)$ for a diffusive sample of thickness $L$ at wavelength $\lambda$. Various schemes have been developed recently to increase the range of the angular memory effect, e.g., by time gating~\cite{2018_Judkewitz_Optica}, spatial filtering~\cite{2019_Chen_OL} of the transmitted light or disorder engineering~\cite{2018_Jang_NatPhoton}, as well as by combining it with the translational memory effect~\cite{2017_Vellekoop_Optica, 2015_Yang_NatPhys, 2018_Arruda_PRA} through a forward-scattering medium, or by coupling light into high-transmission eigenchannels in a diffusive medium~\cite{2019_Yilmaz_PRL}. What all these works have in common, however, is that they are constrained by the restrictions of  the conventional memory effect for which the output wavefront tilts by the same angle and in the same direction as the input wavefront. To overcome these inherent limitations, we consider here a radical expansion of the angular memory effect by addressing the following questions: Is it possible to achieve a tilt in the output wavefront along a different direction and/or with a different angle as compared to that of the input wavefront? Can ``perfect correlation'' be obtained at arbitrarily chosen tilt angles which will effectively increase the memory effect range well beyond the conventional one?

The affirmative answers we provide here to these questions involve a customization of the angular memory by shaping the incident wavefront of a coherent beam. For this purpose, we introduce an angular memory operator whose eigenvectors provide perfect memory for arbitrarily chosen input and output tilt angles of the incoming and outgoing wavefronts. By launching coherent light through such an eigenvector into a diffusive system, we experimentally demonstrate that the transmitted wavefront can even tilt in the opposite direction to that of the incident wavefront. Moreover, we show that such correlations can be achieved simultaneously at different input and output angles and that the corresponding tilt angles at both the input and output may well exceed the conventional angular memory range. Our methodology is applicable to other types of memory effects and in different kinds of complex systems such as chaotic cavities or multimode fibers. It provides a general framework for designing and creating desired memories for various applications in imaging, metrology, and communication through complex media.

\begin{figure*}[ht]
\centering
\includegraphics[width=\linewidth]{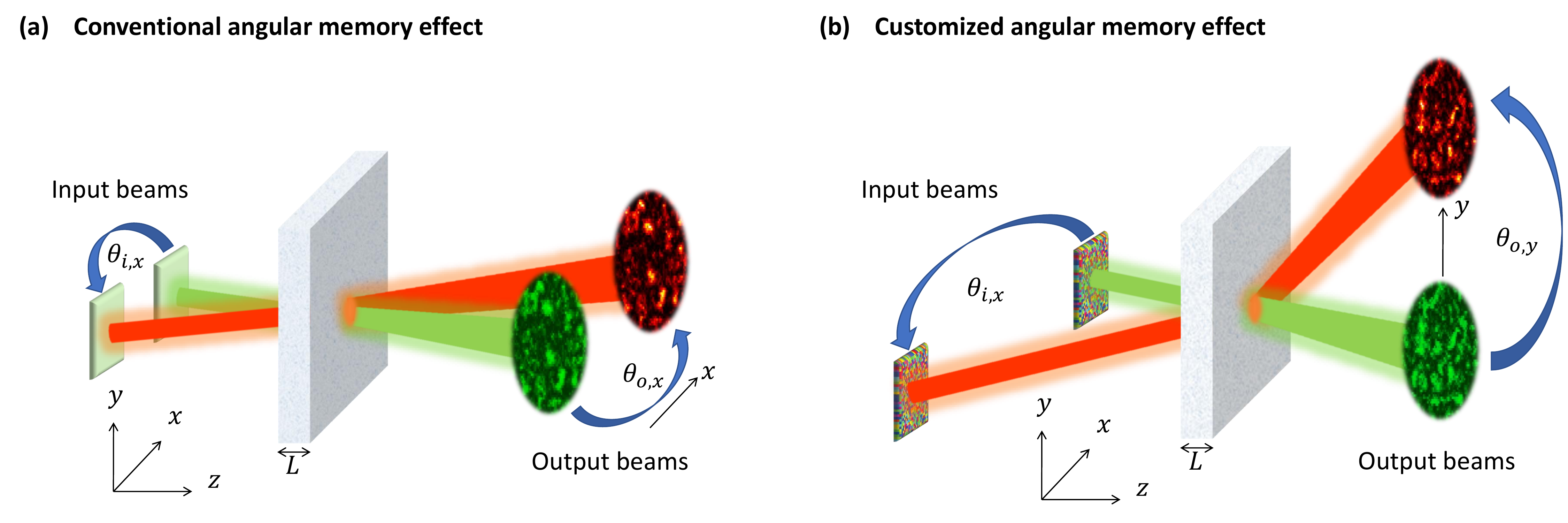}
\caption{\textbf{Conventional vs.\ customized angular memory effect.} \textbf{(a)} Schematic of the conventional angular memory effect: a beam of coherent light (wavelength $\lambda$) is impinging onto a diffusive slab of thickness $L$. When the incident beam is tilted by a small angle $\theta_{i, x} < \lambda/(2\pi L)$, the transmitted pattern tilts in the same direction by $\theta_{o, x} = \theta_{i, x}$. 
\textbf{(b)} Schematic of the customized angular memory effect: both the input and the output tilt angles can be chosen arbitrarily for a special incident wavefront. In this example, when the input wavefront is tilted by $\theta_{i, x}$ in $\hat{x}$ direction, the output wavefront tilts in $\hat{y}$ direction by $|\theta_{o, y}| \neq |\theta_{i,x}|$.}
\label{figure1}
\end{figure*}

\begin{figure}[ht]
\centering
\includegraphics[width=\linewidth]{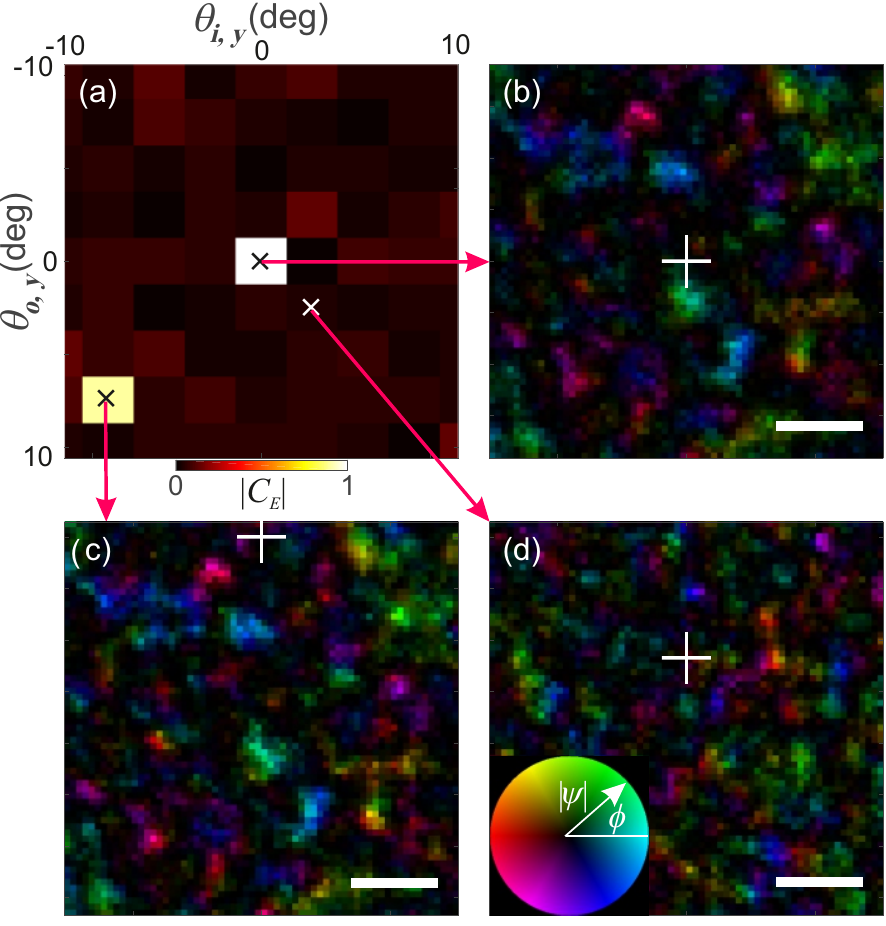}
\caption{\textbf{Angular correlation of output field patterns.} \textbf{(a)} Correlation coefficient $|C_E(\theta_{i,y}, \theta_{o,y})|$ for the incident wavefront given by the eigenvector $V_1$ of the angular memory operator $Q(\tilde{\theta}_{i,y} = -7.8^{\circ}, \tilde{\theta}_{o,y} = 7.1^{\circ})$ with the largest eigenvalue amplitude $|\alpha_1|$. The number of controlled input channels is  $N_i$ = 1024, and the number of detected output channels is $N_o$ = 4096.
\textbf{(b)} With the incident wavefront $V_1$, the transmitted field pattern in the far field is represented with the field amplitude by brightness and the phase by color using the color wheel in \textbf{(d)}. The horizontal and vertical axes denote $k_{o,x}$ and $k_{o,y}$, the components of the output wavevector in $x$ and $y$ directions, respectively. The white plus sign marks the origin: $k_{o,x} = 0$, $k_{o,y} = 0$. The scale bar represents $k = 0.05 (2\pi / \lambda)$.     
\textbf{(c)} When the input wavefront tilts by $\theta_{i,y} = \tilde{\theta}_{i,y} = -7.8^{\circ}$ and the output by $\theta_{o,y} = \tilde{\theta}_{o,y} = 7.1^{\circ}$, the output field pattern is highly correlated to the original one in \textbf{(b)}, with $|C_E| = 0.9$. 
\textbf{(d)} The transmitted field pattern for $\theta_{i,y} = 2.6^{\circ} \neq \tilde{\theta}_{i,y}$ and $\theta_{o,y} = 2.4^{\circ} \neq \tilde{\theta}_{o,y}$ is uncorrelated with that in (b), with $|C_E| = 0.05$. The conventional angular-memory-effect range is about $1^{\circ}$. White plus signs in (c, d) represent the shift of the origin ($k_{o,x} = 0 = k_{o,y}$) of the transmitted field pattern. The scale bars are identical to that in (b). Amplitude and phase patterns in (b,c) are shown in separate panels in Fig.~S2 of the supplementary material.}
\label{figure2}
\end{figure}

\section{Angular memory operator}
To customize the angular memory effect, we first define a correlation coefficient that quantifies the similarity between a transmitted field pattern, $t \ket{\psi}$, and one with arbitrary tilt angles $\theta_i,\theta_o$ of the input and output wavefronts, $X^\dagger({\theta}_o) t X({\theta}_i) \ket{\psi}$:
\begin{equation}
C_E(\theta_i, \theta_o) \equiv \frac{\bra{\psi} t^\dagger X^\dagger(\theta_o) t X(\theta_i) \ket{\psi}}{\sqrt{\bra{\psi}t^\dagger t \ket{\psi} \bra{\psi} X^\dagger(\theta_i) t^\dagger t  X(\theta_i) \ket{\psi}}},
\label{memory_effect}
\end{equation}
where $\psi$ denotes the input field, $t$ the field transmission matrix of the scattering medium, $X(\theta_i)$ and $X^\dagger(\theta_o)$ are rotation operators that tilt the incoming and outgoing field profiles by angles $\theta_i$ and $\theta_o$, respectively. See section~D of the supplementary material for a detailed description of the rotation operator and how to avoid edge effects.

For the conventional angular memory effect the input and the output angles are equal, $\theta_i = \theta_o$, and within the angular range of $\lambda/(2\pi L)$ for a diffusive slab of thickness $L$, as sketched in Fig.~\ref{figure1}(a). To achieve memory for arbitrary $\theta_i$ and $\theta_o$, we tailor the incident wavefront $\psi$ to maximize $|C_E(\theta_i, \theta_o)|$ without compromising the overall transmittance. While nonlinear optimization methods can be employed to search for an optimal $\psi$, they are unlikely to find the global maximum in such a high-dimensional search. We thus introduce here an angular memory operator whose eigenvectors maximize the correlations for any chosen input and output tilt angles. Figure~\ref{figure1}(b) shows, as an example, that while the input wavefront is tilted in the horizontal $(x)$ direction, the output wavefront tilts in the vertical $(y)$ direction, and input and output wavefronts tilt by different angles. 

To build such an angular memory operator, we start with the expression $Q_0 \equiv t^\dagger X^\dagger(\tilde{\theta}_o) t X(\tilde{\theta}_i)$ (appearing in the numerator of Eq.~\ref{memory_effect}), with its eigenvectors given by $Q_0 \ket{V^{(0)}_n} = \alpha^{(0)}_n\ket{V^{(0)}_n}$. In our angular memory operator, $\tilde{\theta}_o$ and $\tilde{\theta}_i$ represent the output and input angles we choose for customizing the angular memory effect, respectively. Using these eigenvectors $\ket{V^{(0)}_n}$ as input wavefronts, the numerator of the correlation coefficient $C_E$ in Eq.~\ref{memory_effect} equals to the corresponding complex-valued eigenvalues $\alpha^{(0)}_n$. Correspondingly, the eigenvectors of $Q_0$ with large $|\alpha^{(0)}_n|$ have a large numerator of $|C_E|$. As it turns out, this does not necessarily enhance $|C_E|$, because the eigenvectors may achieve a large numerator of $|C_E|$ already by coupling light into high-transmission eigenchannels. A higher transmission increases, however, not only the numerator, but also the denominator of $|C_E|$ without necessarily increasing the similarity between $t \ket{V^{(0)}_n}$ and $X^\dagger(\tilde{\theta}_o) t X(\tilde{\theta}_i) \ket{V^{(0)}_n}$. 

To enhance $|C_E|$ instead of just its numerator, we adapt the angular memory operator in the following way: 
\begin{equation}
Q(\tilde{\theta}_i, \tilde{\theta}_o) \equiv (t^\dagger t)^{-1} t^\dagger X^\dagger(\tilde{\theta}_o) t X(\tilde{\theta}_i).
\label{operator}
\end{equation}
The additional term $(t^\dagger t)^{-1}$ counter-balances the increase of the numerator that would result from an increase in the transmittance only. Alternatively, one can also counter-balance the transmittance by using the term $\{[X^\dagger(\tilde{\theta}_o)  tX(\tilde{\theta}_i) ]^\dagger X^\dagger (\tilde{\theta}_o)  tX(\tilde{\theta}_i) \}^{-1}$, which results in a similar expression in Eq.~\ref{operator} (see section~E in the supplementary materials for the derivation). Here, we restrict ourselves to the case where the number of output channels $N_o$ in the transmission matrix $t$ is no less than the number of input channels $N_i$, i.e.\ $N_o \ge N_i$. This situation is typically realized in experimental measurements of a transmission matrix $t$, where the number of controlled input channels $N_i$ (i.e.\  number of columns of $t$) does not exceed the number of detected output channels $N_o$ (i.e.\  number of rows of $t$), and guarantees that the expression $(t^\dagger t)^{-1} t^\dagger$ in Eq.~\ref{operator} (left inverse of $t$) exists. Moreover, when the number of input and output channels is the same, $N_o = N_i$, the left inverse just equals $t^{-1}$, and the angular memory operator in Eq.~\ref{operator} reduces to the simple expression $Q(\tilde{\theta}_i, \tilde{\theta}_o) = t^{-1} X^\dagger(\tilde{\theta}_o) t X(\tilde{\theta}_i)$, with its eigenvectors $Q(\tilde{\theta}_i, \tilde{\theta}_o) \ket{V_n} = \alpha_n\ket{V_n}$ satisfying $X^\dagger(\tilde{\theta}_o) t X(\tilde{\theta}_i) \ket{V_n} = \alpha_n t\ket{V_n}$. More precisely, tilted output field patterns  $X^\dagger(\tilde{\theta}_o) t X(\tilde{\theta}_i) \ket{V_n}$'s are identical to the original ones $t\ket{V_n}$'s, aside from a constant factor $\alpha_n$. The correlation coefficient thus reaches its maximal value, $|C_E| = 1$. Hence, with the number of input and output channels being equal, $N = N_i = N_o$, the $N$ eigenvectors of $Q$, regardless of their associated eigenvalues, provide $N$ incident wavefronts to create $N$ perfectly-correlated pairs of input-output field patterns for the chosen angles $\tilde{\theta}_i$ and $\tilde{\theta}_o$. We note that eigenvectors of a different operator were recently developed to correlate the transmitted field profile of a scattering medium to that of free space~\cite{2021_Pai_Nat_Photon}.

When $N_o>N_i$, the eigenstates of $Q$ satisfy $t^\dagger X^\dagger(\tilde{\theta}_o) t X(\tilde{\theta}_i) \ket{V_n} = \alpha_n t^\dagger t\ket{V_n}$, which is to say that the outputs projected by $t^\dagger$ have perfect correlation. The unprojected outputs can still exhibit high correlations, even though less than unity, and the eigenstate with the highest possible correlation will be shown below.

\section{Experiments}
Next, we construct the angular memory operator $Q$ using our experimentally-measured transmission matrix of a diffusive sample. Our sample is a densely-packed zinc oxide (ZnO) nanoparticle layer on a cover slip. The layer thickness is about 10 \textmu m, much larger than the transport mean free path $l_t \sim$ 1 \textmu m, such that the light transport in the ZnO layer is diffusive. The transmission matrix $t$ is measured with an interferometer setup shown in Fig.~S1 of the supplementary material. A linearly-polarized monochromatic laser beam at wavelength $\lambda$ = 532 nm is split and injected into the two arms of the interferometer. A spatial light modulator (SLM) in the sample arm prepares the phase front of the light field, which is then projected onto the front surface of the ZnO layer. The light transmitted through the sample combines with the reference beam from the other arm. Their interference pattern is recorded by a CCD camera placed in the far field of the sample. The phase front of the output field from the sample is recovered from four interference patterns acquired with varying global phases displayed on the SLM. The reference beam has a flat phase front, allowing the retrieval of the relative phase between output fields at different locations in the far field. This information is critical to the construction of the angular memory operator, which requires measurement of the correlation between fields at different output angles (corresponding to different far-field locations). In contrast to the common-path interferometry method~\cite{2019_Yilmaz_PRL, 2010_Popoff_PRL, 2019_Yilmaz_Nat_Photon}, which measures only the relative phases between different input channels to the same output channel (i.e., only relative phases between the columns of $t$), our method measures the phase differences between both columns and rows of the transmission matrix.

\begin{figure}[H]
\centering
\includegraphics[width=\linewidth]{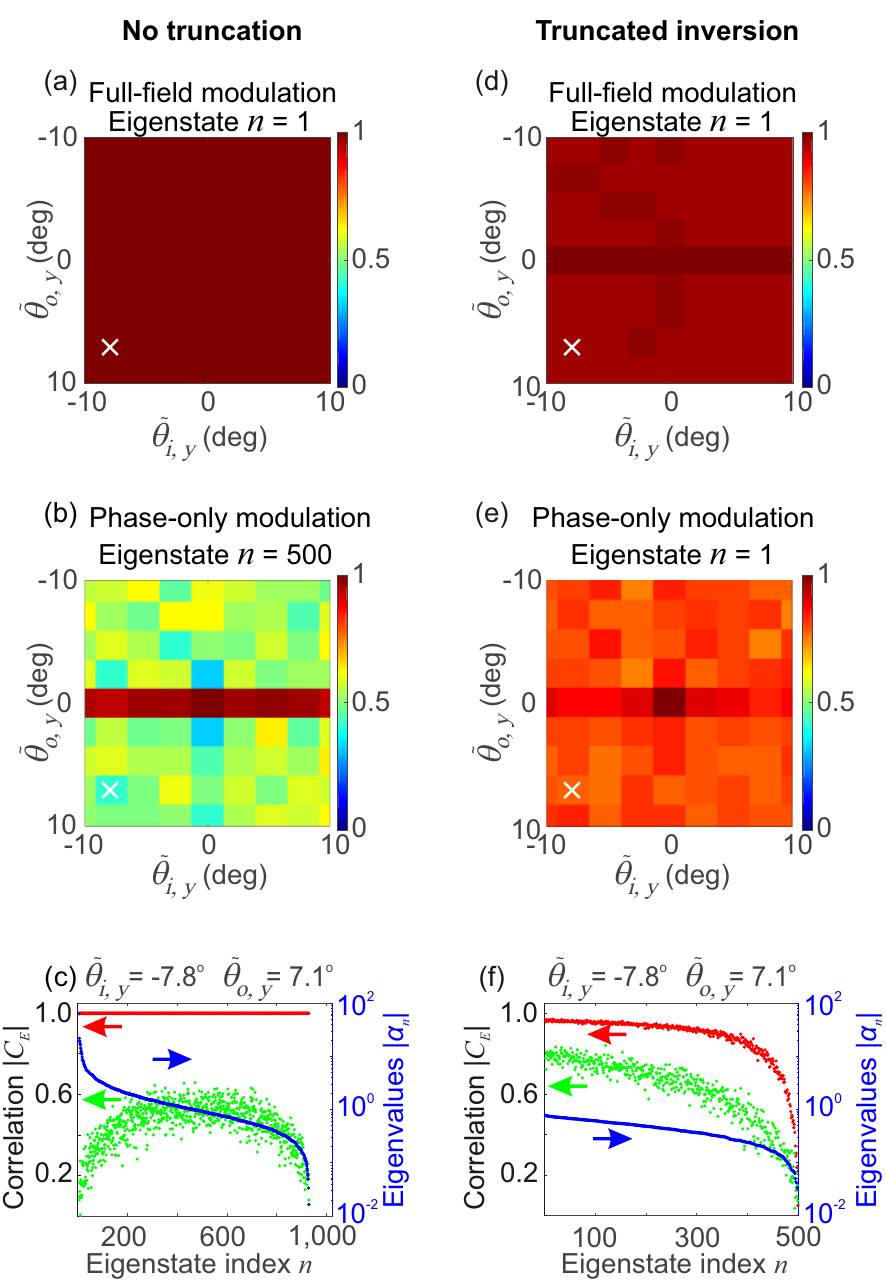}
\caption{\textbf{Angular correlations of eigenstates and eigenvalues of $Q$.} 
\textbf{(a)} All eigenstates of Q customized for any combination of
input and output tilt angles $\tilde{\theta}_i$ and $\tilde{\theta}_o$ have perfect correlation $|C_E| = 1$ at the chosen angles. The number of input/output channels is $N_i = N_o$ = 1024.
\textbf{(b)} Angular correlation $|C_E|$ is reduced when only the phase front of the eigenvector is modulated, except when $\tilde{\theta}_o = 0$. The eigenstate of index $n=500$ has the highest correlation averaged over all input and output tilt angles within the range of $(-10^{\circ}, 10^{\circ})$. 
\textbf{(c)} All 1024 eigenstates of $Q$ for a given pair of $\tilde{\theta}_i = -7.8^{\circ}$ and $\tilde{\theta}_o = 7.1^{\circ}$, marked by white $\times$ in (a,b), have perfect angular correlation $|C_E| = 1$ with full-field modulation of the input wavefront (red, left axis). Phase-only modulation dramatically reduces the angular correlation $|C_E|$ of all eigenstates (green, left axis).
\textbf{(d)} The truncated matrix inversion slightly reduces angular correlation for full-field modulation of the input wavefront, but the first ($n=1$) eigenstate of $Q$ with the highest $|\alpha_1|$ has $|C_E| > 0.95$ for any combination of the input and the output tilt angles $\tilde{\theta}_i$ and $\tilde{\theta}_o$. \textbf{(e)} With phase-only modulation of the input wavefront, the truncated matrix inversion keeps $|C_E| > 0.73$ for the first eigenstate, much improved over that with no truncation in (b). \textbf{(f)} The truncated matrix inversion reduces the number of eigenvalues $\alpha_n$ to 500 (blue, right axis). Their angular correlation with full-field modulation (red, left axis) is compared to that with phase-only modulation (green, left axis). The angular memory eigenstate with a larger eigenvalue amplitude $|\alpha_n|$ has a higher correlation $|C_E|$. White $\times$ signs in (d,e) represent the selected $\tilde{\theta}_i$ and $\tilde{\theta}_o$ for $Q$. In (b,d,e), the eigenstates of $Q$ have varying degree of correlation $|C_E|< 1$, and we select the one with the highest correlation averaged over all pairs of $\tilde{\theta}_i$ and $\tilde{\theta}_o$.}
\label{figure3}
\end{figure}

We construct $Q(\tilde{\theta}_i, \tilde{\theta}_o)$ from the measured $t$. Experimentally we record a part of the total transmission matrix, with the number of input channels $N_i$ equal to the number of SLM macropixels and the number of output channels $N_o$ being determined by the detection area on the camera. Fig.~\ref{figure2} shows an example with $N_i = 1024$ and $N_o = 4096$. Since the camera pixel size is smaller than the average speckle grain size, the number of speckle grains at the output is 455. The transverse plane $x$-$y$ in Fig.~\ref{figure2} is parallel to the sample surface. We choose opposite tilts for the field profiles at the input and the output of the scattering sample, $\tilde{\theta}_{i,y} = -7.8^{\circ}$ and $\tilde{\theta}_{o,y} = 7.1^{\circ}$. This means when the incoming wavefront is tilted in $-y$ direction by $7.8^\circ$, the outgoing wavefront tilts in $+y$ by $7.1^\circ$. We find the eigenvectors of such $Q$, and set the input wavefront to its first eigenvector $\ket{V_1}$ with the largest $|\alpha_1|$. Because $N_i < N_o$, $|C_E|$ does not reach unity, but it increases with $|\alpha_n|$ and has the maximum at $|\alpha_1|$ (see Fig.~S3 in the supplementary material). Then we scan the tilt angle of the input wavefront $\theta_{i,y}$ along the $y$-axis, and calculate the output wavefront using the measured transmission matrix $t$. After tilting the output wavefront by an angle $\theta_{o,y}$ in $y$, we calculate its correlation $C_E$ with the original output wavefront without tilting the input.

\section{Results and discussion}
Fig.~\ref{figure2}(a) shows $|C_E|$ for each pair of $\theta_{i,y}$ and $\theta_{o,y}$ in the range of $-10^{\circ}$ and $10^{\circ}$. The conventional angular memory effect is buried in the correlation at the origin $C_E(0, 0)$, because its range is only about $1^{\circ}$ and is less than the step size ($2.5^{\circ}$) of $\theta_{i,y}$ and $\theta_{o,y}$. Therefore, the chosen angles $\tilde{\theta}_{i,y}$ and $\tilde{\theta}_{o,y}$ are well beyond the conventional angular memory effect range.  Fig.~\ref{figure2}(b) shows the output wavefront without tilting input or output, $\theta_{i,y} = 0$ and $\theta_{o,y} = 0$. When the input wavefront is tilted by $\theta_{i,y} = \tilde{\theta}_{i,y} = -7.8^{\circ}$ and output by $\theta_{o,y} = \tilde{\theta}_{o,y} = 7.1^{\circ}$, the output field pattern in Fig.~\ref{figure2}(c) is very similar to the original pattern in Fig.~\ref{figure2}(b). In contrast, the output field pattern in Fig.~\ref{figure2}(d) is completely different for $\theta_{i,y} = 2.6^{\circ} \neq \tilde{\theta}_{i,y}$ and $\theta_{o,y} = 2.4^{\circ} \neq \tilde{\theta}_{o,y}$. 

In this example (Fig.~\ref{figure2}), the input and the output wavefronts tilt in opposite direction along $y$ axis. In Fig.~S4 of the supplementary material, we present examples where the input wavefront is tilted in $y$ direction, while the output wavefront tilts in $x$ direction or in the diagonal direction $\hat{x} + \hat{y}$. 
As described above, when $N_o \le N_i$, all eigenvectors of $Q(\tilde{\theta}_i, \tilde{\theta}_o)$ should achieve perfect memory with $|C_E(\tilde{\theta}_i, \tilde{\theta}_o)| = 1$ for any input and output tilt angles $\tilde{\theta}_i$ and $\tilde{\theta}_o$.  Fig.~\ref{figure3}(a) shows that such perfect correlation is indeed observed when we reduce $N_o$ to $1024 = N_i$. For any chosen pair of $\tilde{\theta}_i$ and $\tilde{\theta}_o$, the corresponding 1024 eigenstates of $Q$ all have $|C_E| = 1$, as shown in Fig.~\ref{figure3}(c).  

\begin{figure}[H]
\centering
\includegraphics[width=\linewidth]{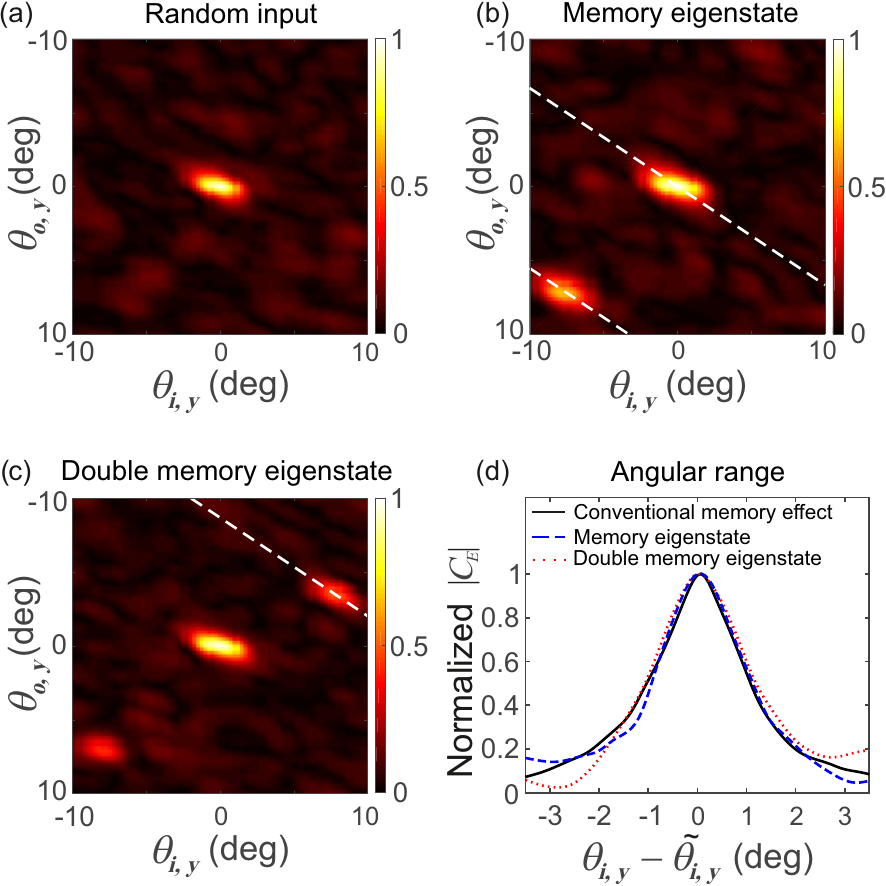}
\caption{\textbf{Experimentally-measured angular memory effect.} 
\textbf{(a)} The angular correlation coefficient $|C_E|$ is enhanced only near the origin for a random input wavefront. $N_i$ = 1024, and $N_o$ = 4096.    
\textbf{(b)} $|C_E|$ increases to 0.66 at ($\theta_{i,y} = -7.8^{\circ}$,  $\theta_{o,y} = 7.1^{\circ}$) when the incident phase front is set to that of the first eigenstate of the angular memory operator $Q$ designed for this set of angles. The conventional memory effect increases the correlation at ($\theta_{i,y} = 0$,  $\theta_{o,y} = 0$).
\textbf{(c)} $|C_E|$ is enhanced at two designated angle pairs far from the origin: $|C_E| = 0.43$ at ($\theta_{i,y} = -7.8^{\circ}$, $\theta_{o,y} = 7.1^{\circ}$), and $|C_E| = 0.5$ at ($\theta_{i,y} = 7.8^{\circ}$, $\theta_{o,y} = -3.5^{\circ}$).
White dashed lines in (b,c) have the slope of $n_i/n_o = 1/1.5$, where $n_i = 1.0$ is the refractive index of air on the input side of the sample and $n_o = 1.5$ is the refractive index of the glass substrate on the output side.
\textbf{(d)} Normalized correlation functions, plotted along the white dashed lines in (b,c), have identical widths for conventional and customized memory effects. The customized memory effects persist over identical angular range around the chosen tilt angles compared to the conventional memory effect around $\theta_i = 0$. In (a-d) we experimentally display phase-only wavefronts for random input and eigenstate inputs. In (b-d) we construct angular memory operators using truncated matrix inversion with the 500 highest transmission eigenchannels.
}
\label{figure4}
\end{figure}

The full memory (perfect correlation) in Figs.~\ref{figure3}(a,c) is obtained from an experimentally measured transmission matrix in case of full-field (amplitude and phase) modulation of the eigenvectors. However, while injecting the eigenvectors experimentally, we use an SLM that modulates only the phase of the input field. When only the phase of an eigenvector is used, however, the correlation $|C_E(\tilde{\theta}_i, \tilde{\theta}_o)|$ drops significantly as shown in Fig.~\ref{figure3}(b). This is because the eigenstates of the angular memory operator consist of both high- and low-transmission eigenchannels, and the latter are strongly affected when the amplitude modulation of the input field is removed. In Fig.~\ref{figure3}(c), all eigenstates of $Q$ have $|C_E|$ significantly less than 1 when only the phase is used. Such a dramatic reduction in $|C_E|$ due to phase-only modulation does not occur in the special case of $\tilde{\theta}_o = 0$, where the angular correlation is encoded in the input wavefront (see Fig.~S8 in the supplementary material for details). To overcome the degradation at $\tilde{\theta}_o \neq 0$, we discard the low-transmission eigenchannels when computing the pseudo-inverse $(t^\dagger t)^{-1} t^\dagger$ in Eq.~(2), a procedure called truncated matrix inversion. In Fig.~\ref{figure3}(d-f), among 1024 transmission eigenchannels, we keep the top 500 with high transmittance. Since the eigenstates of $Q$ comprise only these 500 eigenchannels, their robustness against the absence of amplitude modulation is notably improved. In Fig.~\ref{figure3}(d), the correlation $|C_E|$ with full-field modulation is slightly reduced because the degree of control at the input is reduced by truncated matrix inversion, but the reduction is small. In Fig.~\ref{figure3}(f), we plot $|C_E|$ for all eigenstates of $Q$ with phase-only modulation. The eigenstates with higher $|\alpha_n|$ have stronger correlation, because they have smaller contributions from lower transmission channels, as confirmed by their transmittance shown in Fig.~S5 of the supplementary material. In Fig.~\ref{figure3}(e), the first eigenstate ($n = 1$) with $|\alpha_1|$ closest to 1 has $|C_E|$ above 0.73 for all input and output tilt angles $\tilde{\theta}_i$ and $\tilde{\theta}_o$, even with phase-only modulation of its incident wavefront. 
  
The robustness of the customized angular memory against the absence of amplitude modulation allows us to experimentally excite an eigenvector of the $Q(\tilde{\theta}_i, \tilde{\theta}_o)$ operator with a phase-only SLM, as shown in Fig.~\ref{figure4}. Considering the case in Fig.~\ref{figure4}, we display on the SLM the phase front of the first eigenstate (with the largest $|\alpha_1|$) of $Q$, which is constructed with truncated matrix inversion, keeping the top 500 transmission eigenchannels out of 1024 (see section~H in the supplementary material). To tilt the incident wavefront on the sample, we laterally shift the phase front on the SLM, which is at the Fourier plane of the sample front (input) surface. The step size for input angle scanning is $0.29^{\circ}$, significantly smaller than the conventional angular memory effect range $\delta\theta = 1.7^{\circ}$. For each tilt angle, we measure the transmitted field profile on the CCD camera with four-phase-shift interferometry. Then we tilt the transmitted wavefront and compute its correlation with the original wavefront. For comparison, we display a random phase front on the SLM to measure the conventional angular memory effect.

For a random input wavefront, the conventional angular memory effect manifests itself as a large correlation $|C_E|$ when $\theta_{i, y}$ and $\theta_{o, y}$ are close to 0 in Fig.~\ref{figure4}(a). Note that the memory exists only when the shift in the incident transverse wave vector equals to the shift in the outgoing transverse wave vector. When the two sides of the medium have the same refractive index, the conventional memory effect exists along the diagonal where $\theta_{i, y} = \theta_{o, y}$. In our case, the refractive index of air ($n_i = 1.0$) above the ZnO layer is lower than that of the glass substrate ($n_o = 1.5$) underneath the layer, so the conventional memory effect is tilted from the diagonal, and the white dashed line denotes $\theta_{o, y}/\theta_{i, y} = n_i/n_o = 1/1.5$. In addition, the unequal sampling rate of input and output tilt angles in our experiment contributes to the off-diagonal tilt, as detailed in section~I of the supplementary material.

When the SLM is configured to display the input phase front of the first eigenstate of the $Q(\tilde{\theta}_i, \tilde{\theta}_o)$ operator, $|C_E|$ is greatly enhanced at the preselected angles $\theta_{i,y} = \tilde{\theta}_{i,y} = -7.8^{\circ}$ and $\theta_{o,y} = \tilde{\theta}_{o,y} = 7.1^{\circ}$ of the $Q$ operator in Fig.~\ref{figure4}(b). In spite of the phase-only modulation of input fields and a relatively large ratio $N_o/ N_i = 4$,  $|C_E| = 0.66$ is obtained experimentally. The phase front of the incident eigenstate, displayed on the SLM, does not exhibit any spatial correlation. As shown in Fig.~S8 of the supplementary material, autocorrelations of both the input and the output field patterns give sharply-peaked functions, confirming the angular correlation is not encoded at the input or the output fields. Instead, the angular memory is created via an interplay between the spatial modulation of the incident field and the deterministic scattering of light in the disordered medium. In special cases such as $\tilde{\theta}_{o} = 0$ or $\tilde{\theta}_{i} = 0$, either the input or the output fields of the eigenstates of $Q$ feature periodic  modulations; more details about these special cases are presented in section~K of the supplementary material. 

Finally, to demonstrate the versatility of our approach, we create the angular memory simultaneously for two different input and output tilt angles. In order to realize this with a single incident wavefront, we construct two angular memory operators: $Q_1(\tilde{\theta}_{i,1}, \tilde{\theta}_{o,1}) = (t^\dagger t)^{-1} t^\dagger X^{\dagger}(\tilde{\theta}_{o,1}) t X(\tilde{\theta}_{i,1})$ and $Q_2(\tilde{\theta}_{i,2}, \tilde{\theta}_{o,2}) = (t^\dagger t)^{-1} t^\dagger X^{\dagger}(\tilde{\theta}_{o,2}) t X(\tilde{\theta}_{i,2})$. Then we obtain the joint operator $Q_{1+2} = (Q_1 + Q_2)/\sqrt{2}$ and find its eigenvectors with $Q_{1+2} \ket{V_n} = \alpha_n\ket{V_n}$. The phase-only modulation of the input wavefront, given by the eigenvector with the highest $|\alpha_n|$, enhances the angular correlation $|C_E(\theta_{i}, \theta_{o})|$ at two locations far from the origin in Fig.~\ref{figure4}(c). This means the incident wavefront has two memories: if tilted by $7.8^\circ$ in $-y$ direction, the transmitted wavefront tilts by $7.1^\circ$ in $+y$; however, if the same wavefront is tilted by $7.8^\circ$ in $+y$ at the input, the output wavefront tilts by $3.5^\circ$ in $-y$ instead. The correlation coefficients $|C_E|$ are reduced  roughly by a factor of $\sqrt{2}$, compared to the case of single memory in Fig.~\ref{figure4}(b). Such reduction is less than that of superimposing the eigenvectors of $Q_1$ and $Q_2$ in the incident wavefront, which would reduce the correlation approximately by a factor 2, as shown in section~J of the supplementary material.

The customized memory effect holds not only at the preselected input and output angles, but also over a range around them. This behavior is similar to that of the conventional memory effect. In Fig.~\ref{figure4}(b), with the incident wavefront equal to the first eigenvector of the angular memory operator $Q(\tilde{\theta}_i, \tilde{\theta}_o)$, the correlation remains high as we scan the tilt angle $\theta_i$ of incoming wavefront around $\tilde{\theta}_i$, meaning the outgoing wavefront remains nearly unchanged but tilted away from $\tilde{\theta}_o$. The scanning range of the customized memory effect, given by the angular width of the normalized correlation function in Fig.~\ref{figure4}(d), is identical to that of the conventional memory effect at the origin $(\theta_i = 0, \theta_o = 0)$. Even when we simultaneously create memories at two pairs of input and output angles in Fig.~\ref{figure4}(c), the scanning range of the memory effect at each pair is the same as that of the conventional memory effect, as shown in Fig.~\ref{figure4}(d).

\section{Conclusion}
In conclusion, we have introduced a general angular memory operator $Q(\tilde{\theta}_i, \tilde{\theta}_o)$ to customize the angular memory effect for any tilt angle of the incident and the transmitted wavefronts. As long as the number of detected output channels does not exceed that of controlled input channels, the eigenstates of $Q$ exhibit perfect correlation for arbitrarily and independently chosen input and output tilt angles and directions. Experimentally, we observe strong correlations even when only the phase of the input field is modulated. Furthermore, we simultaneously create memories for different pairs of input and output tilt angles. 

Although our experiment is performed on a diffusive sample with multiple scattering, our method can also be applied to a thin diffuser with forward scattering or to a multimode fiber with random mode mixing. While in the latter case the conventional angular memory effect does not even exist, our approach allows us to create any desired angular memory by launching coherent light into an eigenstate of the angular memory operator. The angular memory operator is applicable not only to classical waves, bu also to quantum waves, opening the door to customizing quantum correlations between entangled photons in complex systems~\cite{2018_Defienne_PRL, 2020_Lib_SA}. Since the memory effects exist in various domains~\cite{2017_Vellekoop_Optica, 2015_Yang_NatPhys, 2018_Arruda_PRA,  amitonova2015rotational, stasio2015light}, the angular memory operator can be extended to the translational memory operator, the rotational memory operator, etc. Our methodology generalizes the memory effects as a versatile and flexible tool for wavefront shaping applications to classical and quantum waves in complex systems.

\section*{Acknowledgments}
We thank Shanti Toenger, Go\"{e}ry Genty, and Arthur Goetschy for useful discussions at the initial stage of the project. 

This work is supported partly by the Office of Naval Research (ONR) under Grant No.\ N00014-20-1-2197, and by the National Science Foundation under Grant No.\ DMR-1905465. M.K.\ and S.R.\ acknowledge support from the European Commission under project NHQWAVE (Grant No.\ MSCA-RISE 691209) and from the Austrian Science Fund (FWF) under project WAVELAND (Grant No.\ P32300).

The authors declare no competing interests.

\bibliography{references}

\end{document}